\documentclass[preprint,aps,amsmath]{revtex4}
\usepackage{bm}
\usepackage[dvips]{graphicx}

\begin{document}

\title{Correlations in a two-dimensional Bose gas with long range interactions}

\author{B. Laikhtman and R. Rapaport}
\affiliation{Racah Institute of Physics, Hebrew University, Jerusalem 91904
Israel}

\begin{abstract}
We study the correlations of two-dimensional dipolar excitons in coupled
quantum wells with a dipole -- dipole repulsive interaction. We show that at
low concentrations, the Bose degeneracy of the excitons is accompanied by
strong multi-particle correlations and the system behaves as a Bose liquid. At
high concentration the particles interaction suppresses quantum coherence and
the system behaves similar to a classical liquid down to a temperature lower
than typical for a Bose gas. We evaluate the interaction energy per particle
and the resulting blue shift of the exciton luminescence that is a direct tool
to measure the correlations. This theory can apply to other systems of bosons
with extended interaction.
\end{abstract}

\maketitle

\begin{figure}
\includegraphics[scale=0.4]{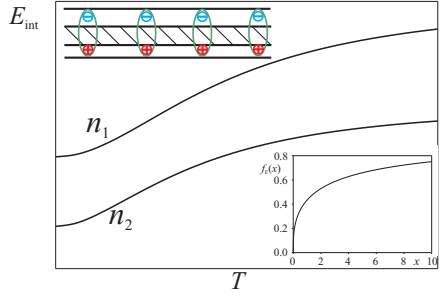}
\caption{\label{fig:E(T)}Typical dependence $E_{int}(T)$ for two different
concentrations $n_{1}>n_{2}$. The upper insert shows the exciton system. In the
lower insert there is function $f_{E}(x)$ that defines the interaction energy
in Eq.(\ref{eq:5}).}
\end{figure}

The investigation of superfluidity of $^{4}$He and Bose condensation of alkali
atoms stimulated the development of the theory of non-ideal Bose gas. The
theory has been developed for contact and hard sphere interactions that
correspond to the hard core repulsion between helium atoms and between alkali
atoms.\cite{Huang,lifshits_sp,Leggett01,Andersen04} During the last few decades
a very intensive search of Bose condensation of excitons in semiconductor
heterostructures resulted in the study of a new type of a Bose system, namely,
a system of indirect excitons in coupled quantum wells in which electrons and
holes are located in different wells, insert in Fig.\ref{fig:E(T)} (see
Refs.\cite{Sonke02,Butov04} and references therein). Such an exciton system is
not only two-dimensional but also the interaction between excitons is
essentially a dipole -- dipole repulsion that extends to a much larger distance
than the exciton radius or the separation between the wells.

The purpose of the present letter is to demonstrate that the relatively large
range of the dipole -- dipole interaction compared to the contact interaction
of bosons results in a dramatic difference of the system behavior. One of the
main new features is that at \textit{any} small concentration the quantum
degeneracy of the gas with the reduction of its temperature is accompanied by a
setting in of multi-particle correlations. As a result the degenerate system
cannot be considered as a weakly non-ideal Bose gas but it is rather a Bose
liquid. The other feature is that at higher concentration the repulsion between
the bosons squeezes the wave function of each one of them and significantly
reduces their overlap. As a result the temperature of setting of a quantum
coherence is reduced compared to what is expected from a Bose gas.

The importance of the long range repulsion has been recently demonstrated by
Zimmermann and Schindler.\cite{Zimmermann07} They calculated the blue shift of
the exciton luminescence line which is essentially the average interaction
energy per one particle. With the help of numerical calculations Zimmermann and
Schindler found a strong pair correlation between excitons that was equivalent
to a depletion region around each one of them. This depletion reduces the
average interaction energy by about 10 times compared to usually used mean
field value! This result is not only fundamental but also has a practical
importance for experimentalists because the measurement of the blue shift is
one of the main ways to extract the exciton concentration.

In general, the reduction of the pair correlation function to zero at small
distance has been noticed long ago\cite{Dingle49} and the crucial new point is
the large length scale of this reduction\cite{Astrakharchik07}.

In this letter we use analytical methods to study the correlation between
excitons in the whole temperature -- concentration plane. In regions where the
system can be considered as gas and only pair correlation is important we
obtain analytical expressions for the pair correlation function and for the average
interaction energy. At low temperatures and high concentrations where the system
behaves as a liquid we use qualitative physical arguments to obtain reasonable
estimates.

The interaction energy between excitons in coupled quantum wells is
\begin{equation}
U(r) = \frac{2e^{2}}{\kappa}
    \left(\frac{1}{r} - \frac{1}{\sqrt{r^{2} + L^{2}}}\right)
\label{eq:1}
\end{equation}
where $L$ is the separation between the centers of the wells and $\kappa$ is
the dielectric constant. According to Ref.\cite{Lozovik96} the attractive Van
der Waals interaction is negligible in practically important values of $L$
($L\gtrsim10$ nm). If any correlation between excitons is neglected then the
average number of excitons in area element $d^{2}\bm{r}$ is $nd^{2}\bm{r}$
where $n$ is the exciton concentration and the average interaction energy is
described by the "plate capacitor formula":
\begin{equation}
E_{int} = \int U(r) n d^{2}r = \frac{4\pi ne^{2}L}{\kappa} \ .
\label{eq:2}
\end{equation}
This result means that the main contribution to $E_{int}$ comes from the
interaction between excitons at distance of the order of $L$: $(e^{2}/\kappa
L)(n\pi L^{2})=\pi ne^{2}L/\kappa$. The same expression is obtained from
quantum mechanical calculation in the mean field
approximation.\cite{Ben-Tabou01,Ivanov02}

A necessary condition for the previous calculation as well as for the
calculations that follow is $nL^{2}\ll1$. In the opposite case the electron --
electron and hole -- hole repulsion is stronger than the electron -- hole
attraction and it is hardly possible to expect a stable exciton phase.
Practically, for excitons in typical coupled quantum well systems, this means
that $n\ll10^{12}$ cm$^{-2}$.

In reality, the correlation between excitons is not small. Two excitons with an
energy of the relative motion $E$ can come to each other only at the distance
larger than $r_{0}(E)$ where $U(r_{0}) = E$. Therefore, Eq.(\ref{eq:2}) is
valid only if the region of strong correlation is small, i.e., $r_{0}\ll L$.
The average exciton energy is of the order of temperature $T$ and the last
condition is equivalent to $T\gg e^{2}/\kappa L$. This condition can also be
written as $\lambda_{T}\ll\pi\sqrt{2bL}$ where
$\lambda_{T}=\pi\hbar\sqrt{2/MT}$ is the thermal wavelength and
$b=\hbar^{2}\kappa/Me^{2}$. The expression for $b$ differs from the Bohr radius
$a_{B}=\hbar^{2}\kappa/me^{2}$ only by replacement of the electron - hole
reduced mass $m=m_{e}m_{h}/(m_{e}+m_{h})$ by the exciton mass $M=m_{e}+m_{h}$.
In GaAs/AlGaAs structures the electron effective mass $m_{e}=0.067$ (in units
of the free electron mass) and the hole effective mass at the bottom of hh1
level in a quantum well $m_{h}=0.14$. This gives $M\approx0.21$ and $b\approx3$
nm which allows us to assume in further calculations that $b\ll L$. Therefore
$\lambda_{T}\ll\pi\sqrt{2bL}$ means also that $\lambda_{T}\ll\pi L$ which
justifies classical description of the exciton -- exciton interaction.

While the validity of Eq.(\ref{eq:2}) requires $T\gg e^{2}/\kappa L\approx140$
K at $L\approx10$nm, the exciton binding energy $\epsilon_{b}$ puts an upper
limit to the exciton temperature. For $L\approx10$ nm $\epsilon_{b}\approx6$
meV = 70 K, i.e., if $T\gg e^{2}/\kappa L$ more than 50\% of excitons are
ionized. There are also significant quantum corrections to $U(r)$ due to
overlap of the exciton wave functions because the exciton radius is not much
smaller than 10 nm.

These estimates demonstrate that the accuracy of Eq.(\ref{eq:2}) is quite
limited.

At $T<e^{2}/\kappa L$ the minimal distance between excitons $r_{0}$ is larger
than $L$ and therefore it becomes the characteristic length scale of the
exciton -- exciton interaction. At this scale Eq.(\ref{eq:1}) can be simplified
and
\begin{equation}
U(r) = \frac{e^{2}L^{2}}{\kappa r^{3}} \ , \hspace{1cm}
    r_{0}(T) = \left(\frac{e^{2}L^{2}}{\kappa T}\right)^{1/3} \ .
\label{eq:3}
\end{equation}
At $r\sim r_{0}$ the exciton correlation is strong and
\begin{equation}
E_{int} = n \int U(r)g(r) d^{2}r
\label{eq:4}
\end{equation}
where $g(r)$ is the pair correlation function.

We now introduce two important dimensionless quantities. the first is the
'interaction parameter': $r_{0}/\lambda_{T}$, which is a measure of the quantum
character of the interaction, and the second is the gas parameter:
$nr_{0}^{2}$, which describes the "diluteness" of the exciton gas. When $T$
becomes smaller than $e^{2}/\kappa L$ the exciton wave length is still smaller
than $r_{0}$, i.e. $r_{0}/\lambda_{T}\ll 1$, and interaction between excitons
can be considered with the help of classical mechanics. In this case
$g(r)=e^{-U(r)/T}$ [see, e.g., Ref.\cite{Hill} Sec.32] that results in
\begin{equation}
E_{int} = \frac{4\pi ne^{2}L}{\kappa} \
    f_{E}\left(\frac{e^{2}}{\kappa LT}\right) ,
\label{eq:5}
\end{equation}
Function $f_{E}(x)$ is shown in the insert in Fig.\ref{fig:E(T)}. In the extreme
case, $T\ll e^{2}/\kappa L$, it becomes:
\begin{equation}
E_{int} = 2\pi\Gamma(4/3)
    n \left(\frac{e^{2}L^{2}}{\kappa}\right)^{2/3}T^{1/3} =
    2\pi\Gamma(4/3)nr_{0}^{2}T \ .
\label{eq:6}
\end{equation}
This result can be understood in the following way. Around each exciton there
is a depletion region of radius $\sim r_{0}$. Without repulsion this region
would contain $nr_{0}^{2}$ excitons with an average energy $T$. The energy
necessary to force all of them out of the region is of the order of
$nr_{0}^{2}T$.

Further reduction of temperature can violate one of the two conditions of
validity of Eq.(\ref{eq:5}). First, $r_{0}/\lambda_{T}\propto T^{1/6}$ can
become smaller than unity, i.e., the exciton wavelength becomes larger than the
potential length scale and the pair correlation has to be described with
quantum mechanics. Second, gas parameter $nr_{0}^{2}\propto1/T^{2/3}$ can
become larger than unity. This makes multi-particle scattering and
multi-particle correlations important, which is characteristic not for a gas
but for a liquid. Which violation happens first depends on the concentration.

If $n\ll(b/2L^{2})^{2}$ then with decrease of temperature the exciton
wavelength $\lambda_{T}$ becomes of the order of and then larger than $r_{0}$
while $r_{0}$ is still much smaller than the average distance between excitons
(i.e., $nr_{0}^{2}\ll 1$). In this case the wave function $\psi(\bm{r})$
describing relative motion of two excitons penetrates under the repulsion
barrier between them and the effective minimal distance is smaller than
$r_{0}$. Given the energy of relative motion $E$, the probability to find one
exciton at the distance $r$ from the other is
$\langle|\psi(\bm{r})|^{2}\rangle_{E}$ where $\langle\dots\rangle_{E}$ means
the average over all directions of the wave vector keeping $E$ constant. For a
non-degenerate exciton gas the probability density for an exciton to have
energy $E$ is $(1/T)e^{-E/T}$. That is
\begin{equation}
g(r) = \frac{1}{T}
    \int_{0}^{\infty} \langle|\psi(\bm{r})|^{2}\rangle_{E} e^{-E/T} dE
\label{eq:7}
\end{equation}
where $\psi(\bm{r})$ has to be normalized in such a way that
$g(r)|_{r\rightarrow\infty}=1$. At high temperatures, when $\lambda_{T}\ll
r_{0}$ it is possible to make use of the $\psi(\bm{r})$ semiclassical
approximation. This leads to $g(r)=e^{-U(r)/T}$ and results in Eq.(\ref{eq:5})
as expected. In the opposite case when $\lambda_{T}\gg r_{0}$ (but still within
the classical statistics limit, $\lambda_{T}\ll n^{-1/2}$) $\psi(\bm{r})$ falls
exponentially at $r<L^{2}/b$ and is a slow function of $r$ at $r>L^{2}/b$. The
interaction potential falls like $U(r)\propto1/r^{3}$ and therefore the main
contribution to the integral in Eq.(\ref{eq:4}) comes from the region $r\sim
L^{2}/b$. The exact solution of the Schr\"odinger equation in this region gives
\begin{equation}
\psi(r) = - \frac{2}{\ln(kL^{2}/b)} \
    K_{0}\left(\frac{2L}{\sqrt{br}}\right) ,
\label{eq:8}
\end{equation}
where $k=\sqrt{ME}/\hbar$. The logarithmic dependence of this wave function on
the energy is very weak and with logarithmic accuracy it is possible to put
$k\approx1/\lambda_{T}$. Then
\begin{equation}
g(r) = \frac{4n}{\ln^{2}(L^{2}/b\lambda_{T})} \
    K_{0}^{2}\left(\frac{2L}{\sqrt{br}}\right) , \hspace{1cm} r \ll \lambda_{T} \ ,
\label{eq:9}
\end{equation}
and substitution of Eq.(\ref{eq:9}) in Eq.(\ref{eq:4}) results in
\begin{equation}
E_{int} = \frac{2\pi\hbar^{2}n}{M\ln^{2}(L^{2}/b\lambda_{T})} \ .
\label{eq:10}
\end{equation}

It is important to note that in all cases considered so far $E_{int}$ is larger
than the interaction energy at an average distance between excitons. This means
that the main contribution to $E_{int}$ comes from rear fluctuations of the
interaction energy and these fluctuations are of the order or larger than the
average value. It is intriguing that while at higher temperatures, where the
interaction is classical, there is a clear temperature dependence of $E_{int}$,
as can be seen from Eq.(\ref{eq:5}), at lower temperatures, quantum
interactions yield an almost $T$ independent $E_{int}$, quite similar to the
mean field result, but with a different coefficient.

The gas parameter in the above quantum case is different from classical
$nr_{0}^{2}$ one. The exciton -- exciton scattering crosssection is
\begin{equation}
\sigma = \frac{\pi}{2k\ln^{2}(kL^{2}/b)}
\label{eq:11}
\end{equation}
(in 2D case it has units of length). In other words, $\sigma$ is of the order
of the wavelength in spite the fact that the length scale of the potential
$L^{2}/b$ is smaller than the wavelength (compare Ref.\cite{landau_qm}, Problem
7 to Sec.132). Respectively the mean free path of excitons is
$l\sim\ln^{2}(L^{2}/b\lambda_{T})/n\lambda_{T}$. The gas condition in the
quantum case corresponds to the absence of correlation between different
scattering events which means that the wavelength has to be much smaller than
the mean free path, i.e., $n\ll\ln^{2}(L^{2}/b\lambda_{T})/\lambda_{T}^{2}$.
This inequality is identical with $E_{int}/T\ll1$ and also with the condition
of non-degeneracy with the accuracy of the logarithmic correction. Practically,
the logarithm is not large compared to unity and hence the gas state of the
exciton system is non-degenerate \textit{while the degeneracy is accompanied by
strong interactions and multi-particle correlations between the excitons}. This
is one noticeable outcome of the above results.

On the other hand, if $n\gg(b/2L^{2})^{2}$ then a decrease of the temperature
leads to a violation of the condition $nr_{0}^{2}\ll1$ while
$r_{0}\gg\lambda_{T}$. That is, quantum corrections are negligible and the
system behaves as a classical liquid. The dimensional analysis gives
\begin{equation}
E_{int} = \frac{e^{2}L^{2}}{\kappa} \ n^{3/2}
    f\left(\frac{e^{2}L^{2}}{\kappa T} \ n^{3/2}\right) \ .
\label{eq:12}
\end{equation}
where according to Eq.(\ref{eq:6}) $f(x)=2\pi\Gamma(4/3)x^{-1/3}$ at $x\ll1$.

Now with a further growth of the concentration or a reduction of the
temperature $nr_{0}^{2}$ comes close to unity and a free motion of excitons
between collisions becomes impossible because each of them is confined in
between its neighbors. This classical picture is valid as long as the size of
the confinement region is much larger than the exciton thermal wavelength. In a
quantum language this situation corresponds to an excited state of the exciton
in a well formed by its neighbors. The size of the well $R\sim n^{-1/2}$ and
the motion is still semi-classical as long as $\lambda_{T}\ll R$. The energy at
the bottom of the well $\sim ze^{2}L^{2}/R^{3}$ ($z$ is the number of nearest
neighbors) is of the same order as the depth of the well. Wells for different
excitons are different, they are not static and sometimes some excitons
overcome or tunnel across the surrounding barrier. But at $nr_{0}^{2}>1$ these
rear occasions don't affect the order of magnitude of the estimates. In
general, this picture is similar to a simple classical liquid and formation of
the wells is the starting point of formation of a short range order
characteristic for liquids\cite{March}. Further reduction of temperature brings
particles to lower levels in the wells and makes the wells more stable.
Stronger confinement of the wave functions of each exciton reduces their
overlap. At $nr_{0}^{2}>1$ the energy of the bottom of the well is larger than
temperature and it is the main part of the interaction energy. Eventually at
$nr_{0}^{2}\gg1$ most of the particles are at the ground state in their
corresponding wells and rough estimates can be done assuming that a short range
order has been formed and each exciton is inside a unit cell of a crystal. (We
emphasize that we mean formation of short range order typical in
liquids\cite{March} but not crystallization and formation of long range
order\cite{Lozovik98}.) Then the potential energy of an exciton that is
displaced by small distance $\bm{r}$ from the center of the unit cell is
\begin{equation}
U_{cr}(\bm{r}) \approx \frac{e^{2}L^{2}}{\kappa R^{3}}
    \left(C_{1} + C_{2} \ \frac{r^{2}}{R^{2}}\right) ,
\label{eq:13}
\end{equation}
and $R=[n\sin(2\pi/z)]^{-1/2}$ is the radius of the first coordinate circle.
The values of the constants can be estimated assuming a crystal order within
first few coordinate circles and random distribution outside. The dependence of
the correlation radius on the order appears quite weak and for hexagonal unit
cell $C_{1}\approx10.5$, $C_{2}\approx15$ (compare Ref.\cite{Rapaport07}). If
$T\sim(e^{2}L^{2}/\kappa)n^{3/2}$ (i.e., $nr_{0}^{2}\sim1$) then $r\sim R$ and
there is no short range order in the system. However if $T
\ll(e^{2}L^{2}/\kappa)n^{3/2}$ then the short range order does exist and $r\ll
R$. The energy of the ground state in the potential (\ref{eq:13}) above its
bottom is $\sim\sqrt{2C_{2}(\hbar^{2}/MR^{2})(e^{2}L^{2}/\kappa R^{3})}$ and
quantization is important when temperature is of the order or smaller than this
value, i.e., $T\lesssim(\hbar^{2}n/M)\sqrt{2C_{2}(L^{2}n^{1/2}/b)}$. The order
or magnitude of $r$ is controlled by the temperature or quantization and in any
case when the short range order does exist the size of a single exciton wave
function is much smaller than the distance between excitons.

The inequality $r\ll R$ allows us to make two conclusions. First, the bottom of
potential of Eq.(\ref{eq:13}) gives a good estimate for the interaction energy
$E_{int}\approx8e^{2}L^{2}n^{3/2}/\kappa$. Second, it is possible to estimate
the overlap of the wave functions of adjacent excitons. If $L=10$ nm and
$n=3\times10^{11}$ cm$^{-2}$ the estimate according to the wave function in the
harmonic potential of Eq.(\ref{eq:13}) gives for the overlap $\sim0.2$. The
actual value is even smaller because Eq.(\ref{eq:13}) is valid only for small
values of $r$ and the potential barrier at larger $r$ is actually steeper. Due
to the small wave function overlap the temperature at which the phase coherence
or/and spin coherence\cite{Fernandez-Rossier96} in the exciton system is set in
is reduced compared to its the expected value $\sim\hbar^{2}n/M$. This is
interesting as it points to a non-monotonic dependence of the quantum coherence
onset temperature on the concentration, and is suggests that a lower density
exciton system can become quantum coherent at higher temperatures than a higher
density system, which is a-priori non intuitive.

\begin{figure}
\includegraphics[scale=0.35]{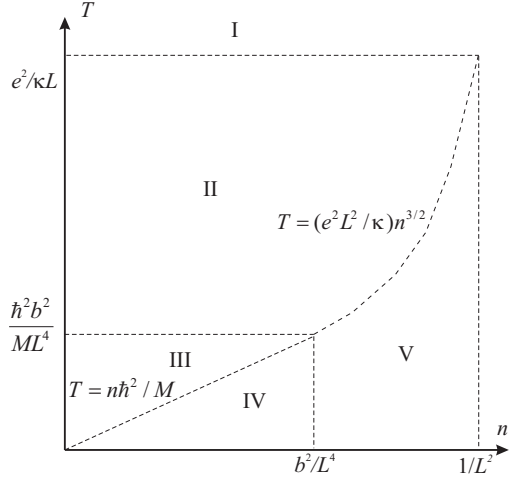}
\caption{\label{fig:nT}Different role of exciton - exciton correlation in the
exciton system. In region I correlation is not important. II is the region of
classical gas with strong pair correlation. In the region III the system is
non-degenerate gas with quantum exciton - exciton scattering and strong pair
correlation. IV is the region of quantum liquid. V is the region of classical
liquid down to very low temperature.}
\end{figure}

All estimates made above open the possibility to make up a general picture that
demonstrates the role of particle correlations in a dipolar exciton system at
the whole $n-T$ plane. This picture is presented in Fig. \ref{fig:nT}.
Correlation is not important and the mean field approximation is applicable
only in region I. In region II the exciton system can be considered as a
classical gas with strong pair correlations. The difference between region III
and II is that in the former exciton - exciton scattering is described by
quantum mechanics that changes the pair correlation function. Reduction of the
temperature from region III to region IV leads to degeneracy of the exciton
system. But simultaneously a strong many - particle correlation is set up. The
system cannot be considered as gas, and the mean free path does not exist.
Rather surprising is the existence of region V where the system behaves as a
classical liquid down to temperatures much lower than the degeneracy
temperature in the case of contact interaction $\sim\hbar^{2}n/M$. The reason
is that strong repulsion between excitons squeezes the wave function of each
exciton to a size smaller than the average distance between excitons. In this
region there is short range order and with a reduction of temperature the
correlation radius grows, i.e., the system may crystallize. However, contrary
to regular classical liquids, the attractive part of the exciton - exciton
interaction is negligible \cite{Lozovik96} and it is likely that if a long
range order is settled it is not as a result of a phase transition but as
gradual growth of the correlation radius.

It is necessary to note that the lines separating the different regions in
Fig.\ref{fig:nT} do not correspond to any sharp transition. Crossing of one of
the lines by changing temperature or concentration leads to a gradual change of
the correlation between excitons. The study of the phase diagram is beyond the
scope of this paper.

The exciton interaction energy that produces the luminescence blue shift has
different value and different dependence on $T$ or $n$ or both of them in
different regions. This difference makes the blue shift a sensitive tool for
measuring correlations in such an exciton system.

We want to emphasize once again that the existence of region V is the result of
the significant tail of the exciton - exciton repulsion potential. In case of a
short range potential (e.g., hard circles) an overlap of particle wave
functions competes with the repulsion and the region of classical behavior does
not exist. This latter case applies to excitons in one well where there is no
dipole - dipole repulsion and an increase of the concentration leads to a Mott
transition but not to a classical liquid.

In conclusion, 2D exciton gas with dipole -- dipole repulsion never can be
considered as degenerate Bose gas with weak interaction. At high concentration
the repulsion suppresses all quantum effects. We evaluated the luminescence
blue shift which is a sensitive tool for measuring of correlations in the
system.

B.L. appreciates discussions with M. Stern and A. L. Efros.


\begin{thebibliography}{99}

\bibitem{Huang}K. Huang, {\it Statistical Mechanics} (John Wiley \& Sons, New
York, 1987).

\bibitem{lifshits_sp}E. M. Lifshitz and L. P. Pitaevskii,
{\it Statistical Physics}, pt.2 (Pergamon Press, Oxford, New York, 1980).

\bibitem{Leggett01}A. J. Leggett, Rev. Mod. Phys. {\bf 73}, 307 (2001).

\bibitem{Andersen04}J. O. Andersen, Rev. Mod. Phys. {\bf 76}, 599 (2004).

\bibitem{Sonke02}D. Snoke, Science {\bf 298}, 1368 (2002).

\bibitem{Butov04}L. V. Butov, J. Phys. Condens. Matter, {\bf 16}, R1577 (2004).

\bibitem{Zimmermann07}R. Zimmermann and C. Schindler, Solid State Commun.
{\bf 144}, 395 (2007); C. Schindler and R. Zimmermann, Phys. Rev. B {\bf 78},
045313 (2008).

\bibitem{Dingle49}R. B. Dingle, Phil. Mag. {\bf 40}, 573 (1949); R. Jastrow,
Phys. Rev. {\bf 98}, 1479 (1955).

\bibitem{Astrakharchik07}G. E. Astrakharchik, J. Boronat, I. L. Kurbakov, and
Yu. E. Lozovik, Phys. Rev. Lett. {\bf 98}, 060405 (2007).

\bibitem{Lozovik96}Yu. E. Lozovik and O. L. Berman, JETP Lett. {\bf 64}, 573
(1996); JETP {\bf 84}, 1027 (1997).

\bibitem{Ben-Tabou01}S. Ben-Tabou de-Leon and B. Laikhtman, Phys. Rev. B
{\bf 63}, 125306 (2001).

\bibitem{Ivanov02}A. L. Ivanov, Europhys. Lett. {\bf 59}, 586 (2002).

\bibitem{Hill}T. L. Hill, {\it Statistical Mechanics} (Dover Publication, Inc.,
New York 1987).

\bibitem{landau_qm}L.D.Landau and E.M.Lifshits,
{\it Quantum Mechanics: Nonrelativistic Theory}, (Pergamon Press, Oxford, New
York 1991).

\bibitem{March}N. H. March and M. P. Tosi, {\it Introduction to liquid state
physics} (Allied Publishers, 2004).

\bibitem{Lozovik98}Yu. E. Lozovik and O. L. Berman, Phys. Solid State {\bf 40},
1228 (1998).

\bibitem{Rapaport07}R. Rapaport and G. Chen, J. Phys.: Condens. Matter. {\bf
19}, 295207 (2007).

\bibitem{Fernandez-Rossier96}J. Fern\'{a}ndez-Rossier, C. Tejedor, L.
Mu\~{n}oz and L. Vi\~{n}a, Phys. Rev. B {\bf 54}, 11582 (1996); J.
Fern\'{a}ndez-Rossier and C. Tejedor, Phys. Rev. Lett. {\bf 78}, 4809 (1997);
Phys. Stat. Solidi (a) {\bf 164}, 343 (1997).


\end{thebibliography}
\end{document}